\documentclass[journal]{IEEEtran}

\usepackage{xcolor}

\usepackage{amsmath}
\usepackage{amssymb}

\usepackage{bbm}
\usepackage{graphicx}
\usepackage{color}
\usepackage{placeins}
\usepackage{float}
\usepackage{hyperref}
\usepackage{tabularx,colortbl}

\usepackage{tikz}
\usepackage{pgf,tikz}
\usepackage{pgfplotstable}
\usepackage{pgfplots}
\pgfplotsset{compat=1.14}
\pgfplotsset{colormap={unicolormapgreen}{rgb255(0cm)=(0,255,0); rgb255(1cm)=(0,255,0)}}

\newcommand{\vt}[1]{\boldsymbol{#1}} 
\newcommand{\uv}[1]{\hat{\boldsymbol{#1}}} 
\newcommand{\arr}[1]{\mathsf{#1}} 
\newcommand{\mat}[1]{\boldsymbol{\mathsf{#1}}} 

\newcommand{\im}{\mathrm{i}}
\newcommand{\me}{\mathrm{e}}
\newcommand{\dd}{\mathrm{d}}
\newcommand{\ld}[1]{\check{#1}} 
\newcommand{\T}{\mathsf{T}} 

\newcommand{\Ns}{N_\text{s}}
\newcommand{\Nv}{N_\text{v}}
\newcommand{\Nf}{N_\text{f}}
\newcommand{\Nt}{N_\text{t}}

\begin{document}
%
\title{Large Time Step and DC Stable TD-EFIE Discretized with Implicit Runge-Kutta Methods}

\author{Alexandre~D\'ely, \IEEEmembership{Member,~IEEE,}
        Francesco~P.~Andriulli, ~\IEEEmembership{Senior Member,~IEEE,}
        and~Kristof~Cools.
\thanks{This work is supported in part by the French DGA agency in the framework of grants for doctoral theses in cooperation between France and United Kingdom, and in part by the European Research Council (ERC) under the European Union's Horizon 2020 research and innovation program (ERC project 321, grant No. 724846).}
\thanks{A. D\'ely and F. P. Andriulli are with the Department of Electronics and Telecommunications, Politecnico di Torino, Turin 10129, Italy (e-mail: alexandre.dely@polito.it, francesco.andriulli@polito.it).}
\thanks{K. Cools is with the Department of Applied Mathematics, Delft University of Technology, Delft 2628, Netherlands (e-mail: K.Cools@tudelft.nl).}
}

\markboth{IEEE Trans. on Antennas and Propagation, Vol-??, Issue-?, Month, ????}%
{D\'ely \MakeLowercase{\textit{et al.}}: Large Time Step and DC Stable TD-EFIE Discretized with Implicit Runge-Kutta Methods}

\maketitle

\begin{abstract}

The Time Domain-Electric Field Integral Equation (TD-EFIE) and its differentiated version are widely used to simulate the transient scattering of a time dependent electromagnetic field by a Perfect Electrical Conductor (PEC). The time discretization of the TD-EFIE can be achieved by a space-time Galerkin approach or, as it is considered in this contribution, by a convolution quadrature using Implicit Runge-Kutta methods. The solution is then computed using the Marching-On-in-Time (MOT) algorithm.
The differentiated TD-EFIE has two problems: (i) the system matrix suffers from ill-conditioning when the time step increases (low frequency breakdown) and (ii) it suffers from the DC instability, i.e. the formulation allows for the existence of spurious solenoidal currents that grow slowly in the solution.
In this work, we show that (i) and (ii) can be alleviated by leveraging quasi-Helmholtz projectors to separate the Helmholtz components of the induced current and rescale them independently. The efficacy of the approach is demonstrated by numerical examples including benchmarks and real life applications.
\end{abstract}

\textbf{\small{\emph{Index Terms}---Time Domain, Electric Field Integral Equation, Implicit Runge-Kutta, Preconditioning, Low Frequency, DC instability.}}

\section{Introduction}

\IEEEPARstart{T}{he} Time Domain-Electric Field Integral Equation (TD-EFIE) can be used  to model the transient scattering from a Perfect Electric Conductor (PEC) \cite{miller_time-domain_1994}.
Many techniques have been developed to improve the solution of Time Domain Integral Equations (TDIE), not only to decrease the overall computational complexity such as the Plane Wave Time Domain (PWTD) \cite{ergin1999plane,shanker2003fast} or the Hierarchical-FFT (HIL-FFT) \cite{yilmaz2002hierarchical} algorithms, but also to improve the stability using loop-tree decomposition \cite{chen2001integral,pisharody2005robust}, Calder\'on preconditioning \cite{andriulli2009analysis}, quasi-Helmholtz projectors \cite{beghein2015dc}, Combined Field Integral Equation (CFIE) \cite{shanker2000analysis,beghein2013space}
and  to improve the accuracy using higher order spatial basis functions \cite{wildman2004accurate,graglia1997higher},  better temporal basis functions \cite{hu2001new,weile2004novel,van2013design}, and exact evaluations \cite{shanker2009time,shi2011stable}.

The most common procedure to solve the TD-EFIE consists in first discretizing in space using a space Galerkin testing with a set of basis functions that spans the space of surface currents, and then discretizing the system in time which results in a fully discretized system that can be solved by a Marching-On-in-Time (MOT) algorithm \cite{miller_time-domain_1994} which is one of the most used scheme although other valid approaches exist such as, for example, marching-on-in-order strategies \cite{chung2004solution}.
The TD-EFIE is often solved after a time differentiation to get rid of the time integration present in the formulation.
Regarding the time discretization, several strategies exist. Point testing is a very common choice \cite{miller_time-domain_1994}, although  the space-time Galerkin discretization is becoming increasingly popular \cite{pray2014higher}.
Another strategy for the time discretization, however, is based on Implicit Runge-Kutta (IRK) convolution quadrature methods \cite{lubich1993runge,banjai2012runge}. 
Similarly to the IRK methods for solving ordinary differential equations, the IRK methods based convolution quadrature for solving TDIE have good stability properties. The accuracy of the solution over time depends on the order of the Runge-Kutta method used.
Discretizations in IRK convolution quadrature methods leverage on  system matrices which can be computed from  Laplace domain integral operators and thus it is relatively easy to get an IRK-based time domain solver from a  frequency domain code. This contrasts with the state-of-the-art for space-time Galerkin methods, where no exponentially converging quadrature schemes for the computation of the interaction integrals are known.
The differentiated TD-EFIE \cite{chen2012analysis} as well as the TD-CFIE \cite{wang2011implicit} have been successfully solved using this IRK methods based convolution quadrature.

Its advantages notwithstanding, the differentiated TD-EFIE still suffers from two serious problems: (i) the large time step breakdown (time domain low frequency (LF) breakdown), which causes the condition number of the system to grow quadratically with the time step \cite{andriulli2009analysis}, and (ii)   the presence of DC instabilities, which corresponds to the existence of static or linear in time solenoidal currents in the solution of the equation \cite{andriulli2009time}. These currents grow slowly yet exponentially, which results in a completely wrong solution.

This paper addresses the solution of both (i) and (ii). In \cite{beghein2015dc}, the LF breakdown and DC instability are solved for the space-time Galerkin discretization by leveraging quasi-Helmholtz projectors. These projectors enable the decomposition of the current in its Helmholtz components, which then can be differentiated or integrated as appropriate. This procedure leads to a discretization that does not suffer from the LF breakdown and DC instability. It remains effective in the case of multiply connected geometries without needing an expensive detection of global loops.  
It is not trivial to see that these projectors can also apply to the IRK convolution quadrature methods, where frequency domain kernels are evaluated in matrix valued complex frequencies. In this paper the regularization of the IRK convolution quadrature discretization of the TD-EFIE is investigated. It is shown how quasi-Helmholtz projectors can be used to arrive at a regularized scheme free from both (i) and (ii). Numerical results demonstrate the efficacy of the proposed approach.
Very preliminary results from this work have been presented in the conference contribution \cite{dely2018stable}.

This paper is organized as follows: in Section~\ref{sec:Background_and_notations}, the background and notations for the classical TD-EFIE discretized with IRK methods are introduced. In Section~\ref{sec:analysis}, the large time step breakdown and the DC instability are analyzed. In Section~\ref{sec:precond_tdefie}, we introduce the new regularized formulation. Implementation details and the computational cost of the scheme are discussed in section \ref{sec:implementation_details}. In Section~\ref{sec:Numerical_results}, numerical results are presented to illustrate the efficacy and the efficiency of the novel approach.

\section{Background and notations}
\label{sec:Background_and_notations}

\subsection{Time domain EFIE}
We consider a PEC object with a boundary $\Gamma$ in a medium whose permittivity is $\varepsilon_0$, permeability is $\mu_0$ and characteristic impedance is $\eta_0 = (\mu_0 / \varepsilon_0)^{1/2}$. This object is excited by an incident wave whose electric field $\vt{e}^\text{inc}$ induces an electric current density $\vt{j}$ on the surface $\Gamma$. The current density $\vt{j}$ satisfies the TD-EFIE 
\begin{equation}
\label{eq:EFIE_time}
    -\eta_0 \vt{T} \vt{j} (\vt{r}, t) = \uv{n}_{\vt{r}} \times \vt{e}^\text{inc} (\vt{r}, t) \,,
\end{equation}
with the time domain EFIE operator $\vt{T}$ defined as
\begin{subequations}
\begin{align}
\label{eq:operatorT_time}
    \vt{T}\vt{j} (\vt{r}, t) &= -\frac{1}{c_0} \frac{\partial}{\partial t} \vt{T}_\text{s} \vt{j} (\vt{r}, t) + c_0 \int_{-\infty}^t \vt{T}_\text{h} \vt{j} (\vt{r}, t') \dd t' \,,\\
    \vt{T}_\text{s} \vt{j} (\vt{r}, t) &= \uv{n}_{\vt{r}} \times \int_{\Gamma} \frac{ \vt{j}(\vt{r}', t - \frac{|\vt{r}-\vt{r}'|}{c_0}) }{4 \pi |\vt{r}-\vt{r}'|}  \dd S' \,,\\
    \vt{T}_\text{h} \vt{j} (\vt{r}, t) &= \uv{n}_{\vt{r}} \times \nabla \int_{\Gamma} \frac{ \nabla' \cdot \vt{j}(\vt{r}', t - \frac{|\vt{r}-\vt{r}'|}{c_0}) }{4 \pi |\vt{r}-\vt{r}'|} \dd S' \,,
\end{align}
\end{subequations}
where $\uv{n}_{\vt{r}}$ is the normal to $\Gamma$ at ${\vt{r}}$ and $c_0 = (\mu_0 \varepsilon_0)^{-1/2}$  is the speed of light in the medium.
The following differentiated TD-EFIE is also often used to avoid the time integration
\begin{equation}
\label{eq:differentiated_EFIE_time}
    \eta_0 \left(\frac{1}{c_0} \frac{\partial^2}{\partial t^2} \vt{T}_\text{s} \vt{j} (\vt{r}, t) - c_0 \vt{T}_\text{h} \vt{j} (\vt{r}, t)\right) = \uv{n}_{\vt{r}} \times \frac{\partial}{\partial t} \vt{e}^\text{inc} (\vt{r}, t) \,.
\end{equation}
To enforce the uniqueness of the solution, the fields are assumed to vanish in the neighbourhood of $\Gamma$ when $t < 0$.

\subsection{Laplace domain EFIE}

The TD-EFIE is transformed to Laplace domain as
\begin{equation}
\label{eq:EFIE_laplace}
    -\eta_0 \ld{\vt{T}} \ld{\vt{j}} (\vt{r}, s) = \uv{n}_{\vt{r}} \times \ld{\vt{e}}^\text{inc} (\vt{r}, s) \,,
\end{equation}
with the Laplace domain EFIE operator $\ld{\vt{T}}$ defined as
\begin{subequations}
\begin{align}
\label{eq:operatorT_laplace}
    \ld{\vt{T}}\ld{\vt{j}} (\vt{r}, s) &= -\frac{s}{c_0} \ld{\vt{T}}_\text{s} \ld{\vt{j}} (\vt{r}, s) + \frac{c_0}{s} \ld{\vt{T}}_\text{h} \ld{\vt{j}} (\vt{r}, s) \,, \\
    \ld{\vt{T}}_\text{s} \ld{\vt{j}} (\vt{r}, s) &= \uv{n}_{\vt{r}} \times \int_{\Gamma} \frac{\me^{-\frac{s}{c_0}|\vt{r}-\vt{r}'|}}{4 \pi |\vt{r}-\vt{r}'|} \ld{\vt{j}}(\vt{r}', s) \dd S' \,, \\
    \ld{\vt{T}}_\text{h} \ld{\vt{j}} (\vt{r}, s) &= \uv{n}_{\vt{r}} \times \nabla \int_{\Gamma} \frac{\me^{-\frac{s}{c_0}|\vt{r}-\vt{r}'|}}{4 \pi |\vt{r}-\vt{r}'|} \nabla' \cdot \ld{\vt{j}}(\vt{r}', s) \dd S' \,.
\end{align}
\end{subequations}

\subsection{Spacial discretization}

The boundary $\Gamma$ is approximated by a triangular mesh with $\Ns$ edges, $\Nv$ vertices, and $\Nf$ faces. Let $(\vt{f}_m)_{m=1}^{\Ns}$ be the set of $\Ns$ Rao-Wilton-Glisson (RWG) basis functions \cite{rao1982electromagnetic} built on the mesh.
The expressions for the standard Helmholtz subspace bases are simpler when the RWGs are normalized such that $\int_{e_m} \vt{f}_m \cdot \uv{m} \dd l = 1$, where $e_m$ is the edge shared by the two triangles in the support of $\vt{f}_m$ and $\uv{m}$ is the normal to $e_m$, tangent to $\Gamma$ \cite{beghein2015dc}.

The RWG basis functions are used to approximate the surface electric current density
\begin{equation}
    \ld{\vt{j}} (\vt{r}, s) \approx \sum_{n=1}^{\Ns} [\arr{j}(s)]_n \vt{f}_n(\vt{r}) \,,
\end{equation}
where $\arr{j}(s)$ is an array that contains the coefficients of the RWG expansion.

The two sides of the Laplace domain EFIE \eqref{eq:EFIE_laplace} are tested with rotated RWG basis functions $(\uv{n}_{\vt{r}} \times \vt{f}_m)_{m=1}^{\Ns}$, which results in the linear system
\begin{equation}
\label{eq:system_laplace}
    \mat{Z}(s) \arr{j}(s) = \arr{e}(s) \,,
\end{equation}
with
\begin{align}
\label{eq:Zs_mn}
    [\mat{Z}(s)]_{mn} &=  -\eta_0 \int_{\Gamma} \uv{n}_{\vt{r}} \times \vt{f}_m(\vt{r}) \cdot \ld{\vt{T}} \vt{f}_n (\vt{r}, s) \dd S \,, \\
    [\arr{e}(s)]_m &= \int_{\Gamma} \vt{f}_m(\vt{r}) \cdot \ld{\vt{e}}^\text{inc} (\vt{r}, s) \dd S \,.
\end{align}
The computation of these matrix elements is standard (see e.g. \cite{graglia1993numerical}, \cite{sauter2010boundary}).

\subsection{Temporal discretization}
\label{sec:temporal_discretization}

The system \eqref{eq:system_laplace} in the Laplace domain has to be transformed into a discrete time domain system.
To do this, the system is first discretized and then transformed to time domain.
More precisely, the discretization corresponds to converting \eqref{eq:system_laplace} from the Laplace domain to the Z-domain i.e. the discrete counterpart of the Laplace domain.
The IRK methods are used to express the Laplace variable $s$ as a function of the Z-domain variable $z$.
Then, the system is transformed to time domain using the inverse Z transform.
A rigorous justification of the Runge-Kutta convolution quadrature can be found in \cite{lubich1993runge}, \cite{banjai2012runge}. In particular the conditions under which the method results in a stable solution are accurately defined in the references. In short, Runge-Kutta methods based on a correspondence $z(s)$ that maps the left half-plane inside the unit circle will give rise to a marching-on-in-time scheme that is stable in principle. In the presence of quadrature error and/or finite machine precision this property might be violated. The design of a solution method robust under these suboptimal conditions is exactly the subject of this paper.

Using convolution quadrature for the computation of retarded potentials is advantageous because it starts directly from Laplace domain kernels. This means that often highly singular time domain kernels can be avoided. Moreover, large parts of well-established and well-tested frequency domain codes can be reused, including bespoke routines for the integration of spatially singular integrals. The use of Runge-Kutta convolution quadrature in particular has a number of additional advantages. (i) Good stability properties are inherited from the Runge-Kutta methods for solving ordinary differential equations. (ii) The accuracy of the solution over time can be simply improved by increasing the order of the Runge-Kutta method.
Note that in the context of ordinary differential equations, Runge-Kutta methods can be used with an adaptive step size \cite{butcher2016numerical}. This does not generalize to TDIEs since using adaptive steps would break time translation symmetry, and the scheme would not result in a discrete convolutional equation, but a more general system, whose solution requires a much higher computational effort.
It is also worth mentioning that some Runge-Kutta schemes can be derived from (discontinuous) Galerkin methods \cite{lesaint1974finite}.

In the Runge-Kutta method, the time dependent solution $y$ of the ordinary differential equation $\frac{\mathrm{d}y(t)}{\mathrm{d}t} = F(t,y)$ is computed consecutively at $t_i = t_0 + i \Delta t$. The initial condition is known: $y_0 = y(t_0)$. The value of the unknown at the next step is computed by adding to the current value a weighted sum of $p$ interpolants $[\arr{F}_i]_k$ of the slope:
\begin{equation}
    y_{i+1} = y_i + \Delta t \sum_{k=1}^p \arr{b}_k [\arr{F}_i]_k \,,
\end{equation}
where $[\arr{F}_i]_k$ are approximations of intermediate times determined by a fraction of the time step $\arr{c}_i$, evaluated by
\begin{equation}
    [\arr{F}_i]_k = F\left(t_i + \arr{c}_k \Delta t, y_i + \Delta t \sum_{l=1}^p \mat{A}_{kl} [\arr{F}_i]_l\right) \,.
\end{equation}
The method is explicit if each $[\arr{F}_i]_k$ depends only on previous $[\arr{F}_i]_l$ i.e. only if $\mat{A}_{kl} = 0$ for all $l \geq k$. Otherwise the method is implicit. In this work only the Implicit Runge-Kutta (IRK) methods are applicable. This means in particular that the well-known Runge-Kutta 4 method is not admissible.

A Runge-Kutta method with $p$ stages is completely specified by $\mat{A} \in \mathbb{R}^{p \times p}$, $\arr{b} \in \mathbb{R}^{p}$ and $\arr{c} \in \mathbb{R}^{p}$ concisely summarized in its so-called Butcher tableau \cite{butcher2016numerical}
\begin{equation}
\label{eq:butcher_tableau}
{\renewcommand{\arraystretch}{1.2}\begin{tabular}{ c | c }
    $\arr{c}$ & $\mat{A}$ \\ 
    \hline
            & $\arr{b}^\T$
\end{tabular}} \,.
\end{equation}
There exist many classes of Runge-Kutta methods but not all of them are suitable for the discretization of a temporal convolution in the context of time domain integral equations. The matrix $\mat{A}$ must be invertible and must verify $\arr{b}^\T \mat{A}^{-1} \mathbbm{1} = 1$. Lobatto IIIC and Radau IIA fulfill these requirements \cite{banjai2012runge,lubich1993runge}.

The equation after Laplace transform corresponding to the time domain differential equation $\frac{\mathrm{d}y(t)}{\mathrm{d}t} = f(t)$ is $s \ld{y}(s) = \ld{f}(s)$. Upon discretization, it is transformed to a corresponding $\mat{s}(z) \arr{y}(\mat{s}(z)) = \arr{f}(\mat{s}(z))$.
Applying the implicit Runge-Kutta method formally amounts to expressing the Laplace variable $s$ as a matrix valued approximation in terms of $z$ \cite{wang2011implicit}
\begin{equation}
\label{eq:mat_s_function_of_z}
    \mat{s}(z) = \frac{1}{\Delta t} \left( \mat{A} + \frac{\mathbbm{1} \arr{b}^\T }{z-1} \right)^{-1} \,,
\end{equation}
where $\mathbbm{1}$ is a vector of size $p$ that contains only ones.
Applying this same substitution in the Laplace domain integral equation yields
\begin{equation}
\label{eq:system_z}
    \widetilde{\mat{Z}}(z) \widetilde{\arr{j}}(z) = \widetilde{\arr{e}}(z) \,.
\end{equation}
where $\widetilde{\mat{Z}}(z) = \mat{Z}(\mat{s}(z))$.
The evaluation of the EFIE BEM matrix elements for a matrix valued $\mat{s}$ instead of a scalar $s$ can be done by computing the eigenvalue decomposition of $\mat{s}(z)$. In particular, the matrix $\mat{s}$ is diagonalized in the form $\mat{s} = \mat{M} \mat{D} \mat{M}^{-1}$ where $\mat{D}$ is a diagonal matrix that contains the $p$ eigenvalues of $\mat{s}$, and $\mat{M}$ is the matrix that contains the corresponding $p$ eigenvectors in each column. The matrix valued element is then computed by evaluating the BEM matrix element for each scalar eigenvalue of the diagonal and multiplying back $\mat{M}$ and $\mat{M}^{-1}$ on the left and right. Additional details on this procedure can be found in \cite{wang2011implicit}.
We introduce the notations  $\widetilde{\mat{s}} = \mat{I} \otimes \mat{s}$, $\widetilde{\mat{M}} = \mat{I} \otimes \mat{M}$, $\widetilde{\mat{D}} = \mat{I} \otimes \mat{D}$ and $\widetilde{\mat{M}}{}^{-1} = \mat{I} \otimes (\mat{M}^{-1})$ where $\mat{I}$ is the $\Ns \times \Ns$ identity matrix and $\otimes$ is the Kronecker product. From these definitions it follows that
$\widetilde{\mat{s}} =  \widetilde{\mat{M}} \widetilde{\mat{D}} \widetilde{\mat{M}}{}^{-1}$.

The multiplication of the scalar $s$ times a matrix or times a vector   can be thought as the multiplication of the diagonal matrix $s\mat{I}$ times the matrix or the vector. The matrix $s\mat{I}$ has $s$ on each diagonal element so, by replacing $s$ with $\mat{s}$, the matrix $s\mat{I}$ becomes the matrix with $\mat{s}$ on the diagonal, i.e. it becomes $\widetilde{\mat{s}}$.
Explicitly,
\begin{equation}
\widetilde{\mat{Z}}(z) = \mat{Z}(\mat{s}(z)) = \widetilde{\mat{M}}(z) \mat{Z}(\mat{D}(z)) \widetilde{\mat{M}}{}^{-1}(z) \,,
\end{equation}
with
\begin{align}
\label{eq:Tx_matD}
    [\mat{Z}(\mat{D}(z))]_{p(m-1)+k,p(n-1)+k} &= [\mat{Z}(\mat{D}_{kk}(z))]_{mn} \,,
\end{align}
where $\mat{D}_{kk}$ is the $k$-th element on the diagonal $\mat{D}$. Thus the matrix valued $\mat{Z}(\mat{D})$ can be computed from the scalar valued matrices $\mat{Z}(\mat{D}_{kk})$.

Given the Z-transform $X(z)$ of a temporal sequence $(x_n)_{n=0}^{\Nt}$ ($X$ and $x$ can be scalars, vectors or matrices), and a counter clockwise contour C around the origin in the region of convergence of $X(z)$, $x_n$ can be computed using the inverse Z-transform
\begin{equation}
    \label{eq:inverse_z_transform_def}
    x_n = \mathcal{Z}^{-1}(X(z))_n = \frac{1}{2 \im \pi} \oint_C X(z) z^{n-1} \dd z \,.
\end{equation}
When $C$ is chosen to be a circle of radius $\rho$ ($\rho \ne 1$), the integral can be approximated using the trapezoidal rule on a $Q$ subintervals partition of $[-\pi,\pi]$
\begin{subequations}
\begin{align}
    x_n &= \frac{\rho^n}{2 \pi} \int_{-\pi}^{\pi} X(\rho \me^{\im \theta}) \me^{\im \theta n} \dd \theta \\
    \label{eq:trapezoidal_rules}
        &\approx \frac{\rho^n}{Q} \sum_{q=0}^{Q-1}  X(\rho \me^{2 \im \pi \frac{q}{Q}}) \me^{2 \im \pi \frac{q}{Q} n} \,.
\end{align}
\end{subequations}

Performing the inverse Z-transform of the product \eqref{eq:system_z} results in the convolution
\begin{equation}
\label{eq:system_time}
    \sum_{j=0}^{i} \mat{Z}_{j} \arr{j}_{i-j} = \arr{e}_i \mbox{ for each } i \in [0, \Nt] \,,
\end{equation}
where
\begin{subequations}
\begin{equation}
\vt{j} (\vt{r}, (i+\arr{c}_k-1) \Delta t) \approx \sum_{m=1}^{\Ns} [\arr{j}_i]_{p(m-1)+k} \vt{f}_m(\vt{r}) \,, \end{equation}
\begin{equation}
\label{eq:en_t}
    [\arr{e}_i]_{p(m-1)+k} = \int_{\Gamma} \vt{f}_m(\vt{r}) \cdot \vt{e}^\text{inc} (\vt{r}, (i+\arr{c}_k-1) \Delta t) \dd S \,,
\end{equation}
\begin{equation}
    \mat{Z}_j = \mathcal{Z}^{-1}(\mat{Z}(\mat{s}(z)))_j \,.
\end{equation}
\end{subequations}
The coefficients $\arr{c}_k$ ($1 \le k \le p$) are given by the Butcher tableau \eqref{eq:butcher_tableau} and correspond to the fraction of the time step where the $p$ stages occur.

The time discretized system \eqref{eq:system_time} can be rewritten to be solved in $\arr{j}_i$ for each $i \in [0, \Nt]$, which corresponds to the MOT algorithm
\begin{equation}
\label{eq:mot}
    \mat{Z}_0 \arr{j}_i = \arr{e}_i - \sum_{j=1}^{i} \mat{Z}_{j} \arr{j}_{i-j} \,.
\end{equation}
Note that classically, the differentiated TD-EFIE is actually solved which means that the system in the Laplace domain \eqref{eq:system_laplace} is multiplied on both sides by $s$.
Introducing an extra differentiation enables getting rid of the time integration of the divergence of the currents in the TD-EFIE or, differently said, it cancels the $1/s$ factor that appears in the EFIE operator \eqref{eq:operatorT_laplace}.
This is done to limit the number of terms in the convolution \eqref{eq:mot} to a certain constant $N_{\text{conv}}$ in the order of $D/(c_0 \Delta t)$ where $D$ is the diameter of the scatterer.
Indeed, it is possible to truncate the convolution after a certain number of terms as the norm of the impedance matrices $\mat{Z}_{j}$ decreases exponentially \cite{wang2011implicit}. The downside of this time differentiation is that, in addition to the constant solenoidal currents responsible of the DC instability, also linear in time solenoidal currents end up in the kernel.

In this paper, we use the quasi-Helmholtz projectors (reviewed in the next section) to address these issues independently by (i) differentiating only the capacitive part of the operator that contains the integration of the divergence of the current to have a fixed number of terms in the convolution, and (ii) integrating only the inductive part that contains the derivative of the current responsible of the DC instability. Overall these two changes balance the TD-EFIE so that the condition number remains stable at low frequency.

\subsection{Quasi-Helmholtz projectors}

In this subsection the quasi-Helmholtz projectors \cite{andriulli2013well} are briefly reviewed. These projectors can be built from $\mat{\Sigma} \in \mathbb{R}^{\Ns \times \Nf}$, the edge-face connectivity matrix of the mesh that discretizes $\Gamma$. The same matrix is also the transformation basis from the basis of stars \cite{vecchi1999loop} to the RWG basis. It is defined by
\begin{equation}
    \mat{\Sigma}_{ef} =
        \left\{ \begin{array}{ll}
    \pm 1 &  \mbox{if edge $e$ is on the boundary of} \\
      & \mbox{face $f$ clockwise/counterclockwise} \\
    0 & \mbox{otherwise}
        \end{array} \right. \,.
\end{equation}
The quasi-Helmholtz projectors $\mat{P}^{\Sigma}$ and $ \mat{P}^{\Lambda H} \in \mathbb{R}^{\Ns \times \Ns}$ are then constructed using the basic properties of the projectors
\begin{subequations}
\begin{align}
    \mat{P}^{\Sigma} &= \mat{\Sigma} \left( \mat{\Sigma}^\T \mat{\Sigma} \right)^{+} \mat{\Sigma}^\T \,, \\
    \mat{P}^{\Lambda H} &= \mat{I} - \mat{P}^{\Sigma} \,.
\end{align}
\end{subequations}
where the $^+$ denotes the Moore-Penrose pseudo inverse and $\mat{I}$ is the identity matrix.

The projector $\mat{P}^{\Sigma}$ projects on the space of non-solenoidal functions  or \emph{stars}, and the complementing projector $\mat{P}^{\Lambda H}$ projects on the space of solenoidal local and global \emph{loops}. Characterizing the loop space as the range of the complementing projector renders the expensive construction of a basis for the global loop space unnecessary \cite{andriulli2013well}.

\section{Conditioning and DC Stability Analysis}
\label{sec:analysis}

Both the standard and time differentiated TD-EFIEs are plagued by ill-conditioning and DC instability issues. In this section, the manifestation of these issues for the IRK convolution quadrature discretization of the TD-EFIE is discussed. The scaling of the condition number in terms of the time step is derived and the regime solutions of the homogenous TD-EFIE are characterized.

\subsection{Large time step ill-conditioning (low-frequency breakdown)}

As the large time step breakdown is a problem related to the time discretization, the starting point of the analysis is the space discretized system in \eqref{eq:system_laplace}  which is continuous in the Laplace domain.
The matrix $\mat{Z}(s)$ in \eqref{eq:Zs_mn} is rewritten to make explicit the contribution of the two parts of the EFIE operator in \eqref{eq:operatorT_laplace}
\begin{equation}
    \mat{Z}(s) = \eta_0 \left(\frac{s}{c_0} \mat{T}_\text{s}(s) + \frac{c_0}{s} \mat{T}_\text{h}(s) \right) \,,
\end{equation}
where
\begin{subequations}
\begin{align}
    [\mat{T}_\text{s}(s)]_{mn} &= \int_{\Gamma} \uv{n}_{\vt{r}} \times \vt{f}_m(\vt{r}) \cdot \ld{\vt{T}}_\text{s}\vt{f}_n (\vt{r}, s) \dd S \,, \\
    [\mat{T}_\text{h}(s)]_{mn} &= -\int_{\Gamma} \uv{n}_{\vt{r}} \times \vt{f}_m(\vt{r}) \cdot \ld{\vt{T}}_\text{h}\vt{f}_n (\vt{r}, s) \dd S \,.
\end{align}
\end{subequations}
The quasi-Helmholtz projectors can be used to make explicit the behavior of the impedance matrix $\mat{Z}(s)$ in a Helmholtz decomposed basis.
By inserting the identity $\mat{I} = \mat{P}^{\Lambda H} + \mat{P}^{\Sigma}$ on the left and the right of $\mat{Z}(s)$, $\mat{Z}(s)$ can be rewritten in a block matrix form that makes clear the different scalings
\begin{equation}
\label{eq:helmoltz_components_Z}
    \mat{Z} = \eta_0
    \begin{pmatrix}
        \mat{P}^{\Lambda H}  & \mat{P}^{\Sigma}
    \end{pmatrix}
    \hspace{-1px}
    \begin{pmatrix}
        \frac{s}{c_0} \mat{T}_\text{s} & \frac{s}{c_0} \mat{T}_s \\
        \frac{s}{c_0} \mat{T}_\text{s} & \frac{s}{c_0} \mat{T}_\text{s} + \frac{c_0}{s} \mat{T}_\text{h}
    \end{pmatrix}
    \hspace{-1px}
    \begin{pmatrix}
        \mat{P}^{\Lambda H}  \\
        \mat{P}^{\Sigma}
    \end{pmatrix} \,,
\end{equation}
where we have used the property $\mat{P}^{\Lambda H} \mat{T}_\text{h}(s) = \mat{T}_\text{h}(s) \mat{P}^{\Lambda H} = \mat{0}$ \cite{andriulli2013well}.
It is clear from \eqref{eq:helmoltz_components_Z} that the matrix is ill-conditioned at low frequency i.e. when $s \to 0$ since the bottom right block (non-solenoidal testings and sources) scales as $1/s$ while the others (solenoidal testings and/or sources) scale as $s$.

Eventually, the system in Laplace domain is transformed into the discrete time domain where the scalings by $s$ become scalings by $\Delta t^{-1}$, and similarly $1/s$ scales proportionally to $\Delta t$  \eqref{eq:mat_s_function_of_z}. This results in a condition number growth for $\mat{Z}_0$ in \eqref{eq:mot} proportional to $\Delta t^2$.

\subsection{DC instability}

It can be easily checked that the solution of the TD-EFIE $\vt{j} (\vt{r}, t)$ in \eqref{eq:EFIE_time} is determined up to a constant solenoidal current $\vt{j}^{\text{cs}} (\vt{r})$ i.e. if $\frac{\partial}{\partial t}\vt{j}^{\text{cs}} (\vt{r}) = \vt{0}$ and $\nabla \cdot \vt{j}^{\text{cs}} (\vt{r}) = 0$ then $\vt{j} (\vt{r}, t) + \vt{j}^{\text{cs}} (\vt{r})$ is also solution of the EFIE in \eqref{eq:EFIE_time}.
This problem is also present in the differentiated TD-EFIE in  \eqref{eq:differentiated_EFIE_time} for which the solution is determined up to a linear in time solenoidal current $\vt{j}^{\text{ls}} (\vt{r},t)$ that verifies  $\frac{\partial^2}{\partial t^2}\vt{j}^{\text{ls}} (\vt{r},t) = \vt{0}$ and $\nabla \cdot \vt{j}^{\text{ls}} (\vt{r},t) = 0$.
This non-uniqueness is problematic. In fact, because of numerical approximations and floating point truncation errors in the discretized equation, the solution current will show a non-physical constant offset which may even grow exponentially for late time steps \cite{andriulli2009time}.

Further characterization of these spurious currents can be done. For the (non-differentiated) TD-EFIE, the equation $\frac{\partial}{\partial t}\vt{j}^{\text{cs}} (\vt{r}, t) = \vt{0}$ is transformed in the Laplace domain (with the initial condition $\vt{j}^{\text{cs}}(\vt{r}, t = 0) = \vt{0}$) as
\begin{equation}
    s \ld{\vt{j}}^{\text{cs}}(\vt{r}, s) = \vt{0} \,.
\end{equation}
After the discretization and after performing the inverse Z-transform, the spurious current $\arr{j}_i^{\text{cs}}$ verifies the following recurrence relation for all $i \in [1, \Nt]$
\begin{equation}
\label{eq:tdefie_dc_current_characterization}
     \sum_{j = 0}^{i} \mathcal{Z}^{-1}(\widetilde{\mat{s}})_i \arr{j}_{i-j}^{\text{cs}} = \arr{0} \,.
\end{equation}
While initializing the sequence with $ \arr{j}_0^{\text{cs}} = \arr{0}$ should result in $\arr{j}_i^{\text{cs}} = \arr{0}$ for all $i$, this is not the case in practice. Indeed, the solution current will eventually contain a non zero constant solenoidal part because of numerical errors.

A simplification of \eqref{eq:tdefie_dc_current_characterization} can be obtained using the property $\arr{b}^\T {\mat{A}}^{-1} \mathbbm{1} = 1$. This property enables rewriting $\mat{s}(z)$ in  \eqref{eq:mat_s_function_of_z} as a finite number of powers of $z$ using the Sherman-Morrison formula
\begin{equation}
\label{eq:s_simplified}
    \mat{s}(z) = \frac{1}{\Delta t} \left( \mat{A}^{-1} - \mat{A}^{-1} \mathbbm{1} \arr{b}^\T \mat{A}^{-1} z^{-1} \right) \,.
\end{equation}
In this case, a multiplication by $z^{-1}$ corresponds to the previous element of the sequence in time domain. So the inverse Z-transform of \eqref{eq:s_simplified} is
\begin{equation}
\label{eq:s_inverse_z_transform}
    \mathcal{Z}^{-1}(\mat{s})_i = \frac{1}{\Delta t} \left( \mat{A}^{-1} \delta_{i,0} - \mat{A}^{-1} \mathbbm{1} \arr{b}^\T \mat{A}^{-1} \delta_{i-1,0} \right) \,.
\end{equation}
where $\delta$ is the Kronecker delta. Inserting \eqref{eq:s_inverse_z_transform} in \eqref{eq:tdefie_dc_current_characterization} results in the following recurrence relation for the spurious current of the (non-differentiated) TD-EFIE
\begin{equation}
\label{eq:recurrence_dc_current_tdefie}
    \arr{j}_{i}^{\text{cs}} = \widetilde{\mathbbm{1}} \widetilde{\arr{b}}^{\T} \widetilde{\mat{A}}{}^{-1} \arr{j}_{i-1}^{\text{cs}} \,,
\end{equation}
where $\widetilde{\mat{A}} = \mat{I} \otimes \mat{A}$ and $\widetilde{\mathbbm{1}} \widetilde{\arr{b}}^{\T} = \mat{I} \otimes  \mathbbm{1} \arr{b}^{\T}$.

A similar characterization can be done for the differentiated TD-EFIE that is more commonly used. In this case, $\frac{\partial^2}{\partial t^2}\vt{j}^{\text{ls}} (\vt{r},t) = \vt{0}$ is transformed in the Laplace domain (with the initial conditions $\vt{j}^{\text{ls}}(\vt{r}, t = 0) = \vt{0}$ and $\frac{\partial}{\partial t}\vt{j}^{\text{ls}}(\vt{r}, t = 0) = \vt{0}$) as
\begin{equation}
    s^2 \ld{\vt{j}}^{\text{ls}}(\vt{r}, s) = \vt{0} \,.
\end{equation}
Again, after the discretization and after performing the inverse Z-transform, the spurious current $\arr{j}_i^{\text{ls}}$ verifies the following recurrence relation for all $i \in [2, \Nt]$
\begin{equation}
\label{eq:diff_tdefie_dc_current_characterization}
     \sum_{j = 0}^{i} \mathcal{Z}^{-1}(\widetilde{\mat{s}}^2)_j \arr{j}_{i-j}^{\text{ls}} = \arr{0} \,.
\end{equation}
As before, using \eqref{eq:s_simplified}, the recurrence equation \eqref{eq:diff_tdefie_dc_current_characterization} on the spurious current of the differentiated TD-EFIE $\arr{j}_i^{\text{ls}}$ can be written as
\begin{align}
\label{eq:recurrence_dc_current_diff_tdefie}
    \arr{j}_{i}^{\text{ls}} &= \widetilde{\mat{A}} \left( \widetilde{\mat{A}}{}^{-1} \widetilde{\mathbbm{1}} \widetilde{\arr{b}}^{\T} +  \widetilde{\mathbbm{1}} \widetilde{\arr{b}}^{\T} \widetilde{\mat{A}}{}^{-1} \right) \widetilde{\mat{A}}{}^{-1} \arr{j}_{i-1}^{\text{ls}} \nonumber \\
    &- \widetilde{\mat{A}}{}^{2} \left( \widetilde{\mat{A}}{}^{-1} \widetilde{\mathbbm{1}} \widetilde{\arr{b}}^{\T} \widetilde{\mat{A}}{}^{-1} \right)^{2} \arr{j}_{i-2}^{\text{ls}} \,.
\end{align}
Note that both \eqref{eq:recurrence_dc_current_tdefie} and \eqref{eq:recurrence_dc_current_diff_tdefie} are independent of the time step $\Delta t$. 

\section{The Regularized TD-EFIE}
\label{sec:precond_tdefie}

In this section, the semi-discrete TD-EFIE (discrete in space, continuous in time) is regularized by a judicious splitting, scaling, and recombination procedure. The resulting semi-discrete equation is then discretized in time by application of the IRK convolution quadrature method. It is argued that the resulting scheme does not suffer from either large time step breakdown or DC instabilities.

\subsection{Regularization in the Laplace domain}

To regularize the TD-EFIE, the solenoidal part ($\Lambda H$) must be multiplied by $c_0/s$ on the left which corresponds to an integration in time domain, and the non-solenoidal part ($\Sigma$) must be multiplied by $s/c_0$ on the right which corresponds to a derivative. We also introduce a length $a$ to keep a consistent dimensionality. The specific choice of $a$ does not affect the asymptotic behaviour of the preconditioning; it can be safely chosen to be the diameter of the scatterer.
The regularized system is thus 
\begin{align}
    \mat{Z}^{\mbox{reg}} &= \left( \frac{c_0}{sa} \mat{P}^{\Lambda H}  + \mat{P}^{\Sigma} \right) \mat{Z} \left( \mat{P}^{\Lambda H}  + \frac{sa}{c_0} \mat{P}^{\Sigma} \right) \\ 
    &= \eta_0
    \begin{pmatrix}
        \mat{P}^{\Lambda H}  & \mat{P}^{\Sigma}
    \end{pmatrix}
    \begin{pmatrix}
        \frac{1}{a} \mat{T}_\text{s} & \frac{s}{c_0} \mat{T}_\text{s} \\
        \frac{s}{c_0} \mat{T}_\text{s} & a\frac{s^2}{c_0^2} \mat{T}_\text{s} + a\mat{T}_\text{h}
    \end{pmatrix}
    \begin{pmatrix}
        \mat{P}^{\Lambda H}  \\
        \mat{P}^{\Sigma}
    \end{pmatrix} \\
    &= \eta_0 \left( \frac{1}{a}\mat{P}^{\Lambda H} \mat{T}_\text{s}  \mat{P}^{\Lambda H} + a \mat{P}^{\Sigma} \mat{T}_\text{h}  \mat{P}^{\Sigma}  + a \frac{s^2}{c_0^2}  \mat{P}^{\Sigma}  \mat{T}_\text{s} \mat{P}^{\Sigma}
    \right. \nonumber \\ &\phantom{+ \eta_0(} \left.
    \label{eq:Zstab_scalar_developed}
    + \frac{s}{c_0} \left( \mat{P}^{\Sigma} \mat{T}_\text{s}  \mat{P}^{\Lambda H} +  \mat{P}^{\Lambda H} \mat{T}_\text{s} \mat{P}^{\Sigma} \right)  \right) \,. 
\end{align}
As the off-diagonal blocks will vanish when $s \to 0$, the regularized matrix $\mat{Z}^{\mbox{reg}}(s)$ tends to a constant at low frequency. In particular, its condition number will remain stable (it actually tends to the condition  number of  $\frac{1}{a} \mat{P}^{\Lambda H} \mat{T}_\text{s} (0) \mat{P}^{\Lambda H} + a \mat{P}^{\Sigma} \mat{T}_\text{h} (0) \mat{P}^{\Sigma}$).
In the Laplace domain, the preconditioned system now reads
\begin{equation}
\label{eq:stable_system_laplace}
\mat{Z}^{\mbox{reg}}(s) \arr{y}(s) = \arr{v}(s) \,.
\end{equation}
The original current $\arr{j}(s)$ can be retrieved accordingly from the auxiliary current $\arr{y}(s)$
\begin{equation}
\label{eq:j_s_function_of_y_s}
    \arr{j}(s) = \left( \mat{P}^{\Lambda H}  + \frac{sa}{c_0} \mat{P}^{\Sigma} \right) \arr{y}(s) \,.
\end{equation}
The new right hand side $\arr{v}(s)$ can also be expressed as a function of the original RHS $\arr{e}(s)$
\begin{equation}
\label{eq:rhs_laplace_precond}
    \arr{v}(s) = \left( \frac{c_0}{sa} \mat{P}^{\Lambda H}  + \mat{P}^{\Sigma} \right) \arr{e}(s) \,.
\end{equation}

\subsection{Time Discretization}

We follow the same procedure for the time discretization using IRK methods as described in  section \ref{sec:temporal_discretization}.
So we first substitute $s$ by $\mat{s}(z)$ in $\mat{Z}^{\mbox{reg}}(s)$ using \eqref{eq:mat_s_function_of_z} and then we perform an inverse Z-transform \eqref{eq:inverse_z_transform_def}.
It can be summarized with the following formula
\begin{equation}
\label{eq:def_Zstab_sequence}
\mat{Z}_{i}^{\mbox{reg}} = \mathcal{Z}^{-1}\left( \mat{Z}^{\mbox{reg}}\left(\mat{s}\left(z\right) \right) \right)_i \,.
\end{equation}
To be compatible with the $p$ stages of the Runge-Kutta methods, the projectors $\mat{P}^{\Lambda H}$ and $ \mat{P}^{\Sigma} \in \mathbb{R}^{\Ns \times \Ns}$ need to be transformed into $\widetilde{\mat{P}}{}^{\Lambda H} = \mat{P}^{\Lambda H} \otimes \mat{I}_p$ and $\widetilde{\mat{P}}{}^{\Sigma} = \mat{P}^{\Sigma} \otimes \mat{I}_p \in \mathbb{R}^{p \Ns \times p \Ns}$  where $\otimes$ denotes the Kronecker product and $\mat{I}_p$ is the $p \times p$ identity matrix.
An intuitive explanation of this transformation is that the projectors are independent of the Laplace variable $s$, so each coefficient in the projectors can be thought as being multiplied by $s^0$. So after the substitution of $s$ by the matrix valued $\mat{s}$, each coefficient is multiplied by $\mat{s}^0$ i.e. the $p \times p$ identity matrix $\mat{I}_p$.
Using the above notations and definitions, and substituting $s$ by $\mat{s}$ in \eqref{eq:Zstab_scalar_developed}, $\mat{Z}^{\mbox{reg}}\left(\mat{s} \right)$ can be written as
\begin{subequations}
\begin{align}
    \mat{Z}^{\mbox{reg}}\left(\mat{s} \right) &=
    \eta_0 \left( \frac{1}{a} \widetilde{\mat{P}}{}^{\Lambda H} \widetilde{\mat{M}} \mat{T}_\text{s}(\mat{D}) \widetilde{\mat{M}}{}^{-1} \widetilde{\mat{P}}{}^{\Lambda H}
\right.  \\ &\phantom{+ \eta_0(} \left.
    + a \widetilde{\mat{P}}{}^{\Sigma} \widetilde{\mat{M}} \mat{T}_\text{h}(\mat{D}) \widetilde{\mat{M}}{}^{-1} \widetilde{\mat{P}}{}^{\Sigma}
\right.  \\ &\phantom{+ \eta_0(} \left.
    + \frac{a}{c_0^2} \widetilde{\mat{P}}{}^{\Sigma}  \widetilde{\mat{M}} \widetilde{\mat{D}}{}^2 \mat{T}_\text{s}(\mat{D}) \widetilde{\mat{M}}{}^{-1} \widetilde{\mat{P}}{}^{\Sigma}
\right.  \\ &\phantom{+ \eta_0(} \left.
    + \frac{1}{c_0}\widetilde{\mat{P}}{}^{\Sigma} \widetilde{\mat{M}} \widetilde{\mat{D}} \mat{T}_\text{s}(\mat{D}) \widetilde{\mat{M}}{}^{-1} \widetilde{\mat{P}}{}^{\Lambda H}
\right.  \\ &\phantom{+ \eta_0(} \left.
\label{eq:Zstab_matrix_developed}
    + \frac{1}{c_0} \widetilde{\mat{P}}{}^{\Lambda H} \widetilde{\mat{M}} \widetilde{\mat{D}} \mat{T}_\text{s}(\mat{D}) \widetilde{\mat{M}}{}^{-1} \widetilde{\mat{P}}{}^{\Sigma} \right) \,.
\end{align}
\end{subequations}
$\mat{Z}_{i}^{\mbox{reg}}$ in \eqref{eq:def_Zstab_sequence} is computed by inserting the above expression for $\mat{Z}^{\mbox{reg}}\left(\mat{s} \right)$ in the formula for the inverse Z-transform \eqref{eq:trapezoidal_rules}.

Regarding the time discretization corresponding to the excitation vector, as the inverse of $s$ in Laplace domain corresponds to a time integration in time domain, a primitive $\vt{E}^\text{inc}$ of the excitation $\vt{e}^\text{inc}$ needs to be computed ($\partial \vt{E}^\text{inc}/ \partial t = \vt{e}^\text{inc}$). This can be done analytically or numerically by applying an integrator based on the same IRK scheme as used elsewhere. The time discretization of the excitation vector \eqref{eq:rhs_laplace_precond} for the preconditioned TD-EFIE becomes 
\begin{equation}
\label{eq:rhs_precond}
    \arr{v}_i = \frac{c_0}{a} \widetilde{\mat{P}}{}^{\Lambda H} \arr{E}_i + \widetilde{\mat{P}}{}^{\Sigma} \arr{e}_i \,,
\end{equation}
where $\arr{e}_i$ was defined in \eqref{eq:en_t} and $\arr{E}_i$ is defined similarly replacing $\vt{e}^\text{inc}$ by one of its primitives $\vt{E}^\text{inc}$ 
\begin{equation}
\label{eq:primitive_en_t}
[\arr{E}_i]_{p(m-1)+k} = \int_{\Gamma} \vt{f}_m(\vt{r}) \cdot \vt{E}^\text{inc} (\vt{r}, (i+\arr{c}_k-1) \Delta t) \dd S \,.
\end{equation}
The choice of the primitive's constant for $\vt{E}^\text{inc}$ does not change the result of $\arr{v}_i$ as it is cancelled by the multiplication with $\widetilde{\mat{P}}{}^{\Lambda H}$. However in practice to avoid numerical cancellations, the primitive constant should be chosen to have $\vt{E}^\text{inc}(t) = \vt{0}$ when $\vt{e}^\text{inc} (t)$ vanishes for $t \to \pm \infty$.

Transforming the stable system in the Laplace domain \eqref{eq:stable_system_laplace} to the discrete time domain results in the following convolution
\begin{equation}
    \sum_{j=0}^{i} \mat{Z}_{j}^{\mbox{reg}} \arr{y}_{i-j} = \arr{v}_i \,.
\end{equation}
This equation is rewritten to make the MOT scheme evident. In addition, as explained at the end of the section \ref{sec:temporal_discretization}, the number of terms in the convolution can be truncated to $N_{\text{conv}}$ terms
\begin{equation}
\label{eq:mot_stable}
    \mat{Z}_0^{\mbox{reg}} \arr{y}_i = \arr{v}_i - \sum_{j=1}^{N_{\text{conv}}} \mat{Z}_{j}^{\mbox{reg}} \arr{y}_{i-j} \,.
\end{equation}

\subsection{Computation of the Electric Current}

After completion of the MOT procedure \eqref{eq:mot_stable}, we need to reconstruct $(\arr{j}_i)_{i=0}^{\Nt}$ from the auxiliary $(\arr{y}_i)_{i=0}^{\Nt}$ using their corresponding relation in the Laplace domain \eqref{eq:j_s_function_of_y_s}.
Performing an inverse Z-transform on the discrete counterpart of \eqref{eq:j_s_function_of_y_s} results in
\begin{equation}
\label{eq:j_n_function_of_y_n_general}
    \arr{j}_i = \widetilde{\mat{P}}{}^{\Lambda H} \arr{y}_i  + \frac{a}{c_0} \widetilde{\mat{P}}{}^{\Sigma}  \sum_{j=0}^i \mathcal{Z}^{-1}(\widetilde{\mat{s}})_j \arr{y}_{i-j} \,,
\end{equation}
where $\widetilde{\mat{s}}(z) = \mat{I} \otimes \mat{s}(z)$.
A simplification of \eqref{eq:j_n_function_of_y_n_general} can be obtained using \eqref{eq:s_simplified} and reads
\begin{equation}
\label{eq:retrieve_original_solution}
    \arr{j}_i =  \widetilde{\mat{P}}{}^{\Lambda H} \arr{y}_i + \frac{a}{c_0 \Delta t} \widetilde{\mat{P}}{}^{\Sigma} \left(\widetilde{\mat{A}}{}^{-1} \arr{y}_i - \widetilde{\mat{A}}{}^{-1} \widetilde{\mathbbm{1}} \widetilde{\arr{b}}^{\T} \widetilde{\mat{A}}^{-1} \arr{y}_{i-1} \right) \,.
\end{equation} 

\section{Implementation details}
\label{sec:implementation_details}

In this section, we describe how the solution of the stable system \eqref{eq:mot_stable} can be implemented in practice: we explain why the use of the quasi-Helmholtz projectors in the preconditioning is compatible with existing fast solvers. Computational complexity and memory usage are discussed.

\subsection{Quasi-Helmholtz projectors}
Although they exhibit a pseudo inverse in their definitions, the projectors $\mat{P}^{\Sigma}$ and $ \mat{P}^{\Lambda H}$ can be multiplied by a vector in linear complexity $O(\Ns)$  by leveraging algebraic multigrid techniques \cite{livne2012lean, andriulli2013well}. These schemes can be applied also in the case of multiscale scatterers.

Consider the multiplication of $\widetilde{\mat{P}}{}^{\Sigma} = \mat{P}^{\Sigma} \otimes \mat{I}_p$ by some vector $\arr{x}$ whose elements are given by $[\arr{x}]_{p(m-1)+k} = [\arr{x}^k]_m$ where the $\arr{x}^k$ are subvectors of $\arr{x}$ elements of the stage $k$ ($1 \leq m \leq \Ns$, $1 \leq k \leq p$).
Then the elements of the product are $[\widetilde{\mat{P}}{}^{\Sigma} \arr{x}]_{p(m-1)+k} = [\mat{P}^{\Sigma} \arr{x}^k]_m$.
Otherwise said, the multiplication of $\widetilde{\mat{P}}{}^{\Sigma}$ can be computed with $p$ multiplications of $\mat{P}^{\Sigma}$ that are linear in complexity.
As a consequence, $\widetilde{\mat{P}}{}^{\Lambda H}$ and $\widetilde{\mat{P}}{}^{\Sigma}$ can also multiply a vector in $O(\Ns)$  operations. The required storage is $O(\Ns)$ for the sparse matrix $\mat{\Sigma}$.

\subsection{Excitation vectors}
The computation of the excitation vectors $\arr{E}_i$ and $\arr{e}_i$ in \eqref{eq:rhs_precond} is linear in complexity, their multiplications by the projectors is also linear.
So the overall cost of computing the preconditioned excitation vectors $\arr{v}_i$ for each $i \in [0,\Nt]$ is $O(\Ns \Nt)$ both in time and memory.

\subsection{Retrieving the solution current}
When $\arr{y}_i$ and $\arr{y}_{i-1}$ are known, $\arr{j}_i$ is computed using \eqref{eq:retrieve_original_solution}. Again the multiplications by the projectors cost $O(\Ns)$.
Then, as $\widetilde{\mat{A}}{}^{-1}$ and $\widetilde{\mat{A}}{}^{-1} \widetilde{\mathbbm{1}} \widetilde{\arr{b}}^{\T} \widetilde{\mat{A}}{}^{-1}$ are block diagonal matrices with ${\mat{A}}^{-1}$ and $\mat{A}^{-1} \mathbbm{1} \arr{b}^\T \mat{A}^{-1}$ blocks in their diagonals, the multiplications by $\widetilde{\mat{A}}{}^{-1}$ and  $\widetilde{\mat{A}}{}^{-1} \widetilde{\mathbbm{1}} \widetilde{\arr{b}}^{\T} \widetilde{\mat{A}}{}^{-1}$ are also linear in complexity because they actually correspond to $\Ns$ multiplications by ${\mat{A}}^{-1}$ and $\mat{A}^{-1} \mathbbm{1} \arr{b}^\T \mat{A}^{-1}$.
This is done for each time step, so overall the retrieving of the original solution $\arr{j}_i$ for each $i \in [0,\Nt]$ costs $O(\Ns \Nt)$  in time and memory.

\subsection{Marching-on-in-time}
Using an iterative solver to solve for $\arr{y}_i$ in \eqref{eq:mot_stable} at each time step, we can assume that it requires a fixed number of iterations $N_{\text{iter}}$. There are also $N_{\text{conv}}$ terms in the convolution in the RHS. So overall, the MOT \eqref{eq:mot_stable} requires $O((N_{\text{iter}} + N_{\text{conv}}) \Nt)$ multiplications of $\mat{Z}_{j}^{\mbox{reg}}$ by a vector.
Note that the number of iterations $N_{\text{iter}}$ is independent of the time step but it still depends on the discretization density.
The number of terms in the convolution in the RHS
$N_{\text{conv}}$ is proportional to $D/(c_0 \Delta t)$ where $D$ is the diameter of the scatterer. In particular, $N_{\text{conv}}$ is low for large time step, however it is unbounded for small time steps. Fast Runge-Kutta convolution quadrature methods for use in this regime have been described in \cite{banjai2014fast}.

It only remains to explain the multiplication $\mat{Z}_{j}^{\mbox{reg}} \arr{x}$ where $\arr{x}$ is a vector.
Using the definitions of $\mat{Z}_{j}^{\mbox{reg}}$ \eqref{eq:def_Zstab_sequence} and the inverse Z-transform \eqref{eq:trapezoidal_rules} we have
\begin{align}
    \mat{Z}_{j}^{\mbox{reg}} \arr{x} &= \frac{\rho^j}{Q} \sum_{q=0}^{Q-1} \me^{2 \im \pi \frac{q}{Q} j} \mat{Z}^{\mbox{reg}}\left(\mat{s}\left( \rho \me^{2 \im \pi \frac{q}{Q}} \right) \right) \arr{x} \,.
\end{align}
$\mat{Z}_{n}^{\mbox{reg}} \arr{x}$ requires $Q$ multiplications of the form $\mat{Z}^{\mbox{reg}}\left(\mat{s}\right) \arr{x}$. So overall it requires $O(Q(N_{\text{iter}} + N_{\text{conv}}) \Nt)$ multiplications of this form. Note that in our case where the temporal sequence is real, it is possible to take advantage of the complex conjugation to avoid half of the multiplications in the inverse Z-transform.

\subsection{Interaction matrix-vector product}
As it can be read from \eqref{eq:Zstab_matrix_developed}, $\mat{Z}^{\mbox{reg}}\left(\mat{s}\right) \arr{x}$ involves multiplications of $\arr{x}$ by $\widetilde{\mat{M}}$, $\widetilde{\mat{D}}$ and $\widetilde{\mat{M}}{}^{-1}$ that have a $O(\Ns)$ complexity like the multiplication by $\widetilde{\mat{A}}{}^{-1}$  because they are block diagonal matrices as explain above. Also the multiplications by the projectors are linear in complexity. It remains the multiplications of $\mat{T}_{x}(\mat{D})$ by a vector ($\mat{T}_{x}$ is either $\mat{T}_\text{s}$ or $\mat{T}_\text{h}$).
Similarly to the multiplications with the projectors as it was explained above, $\arr{x}$ can be subdivided into $p$ subvectors $\arr{x}^k$ such that $[\arr{x}]_{p(m-1)+k} = [\arr{x}^k]_m$. Then using \eqref{eq:Tx_matD}, the elements of the product are $[\mat{T}_{x}(\mat{D}) \arr{x}]_{p(m-1)+k} = [\mat{T}_{x}(\mat{D}_{kk}) \arr{x}^k]_m$. Otherwise said, the product $\mat{T}_{x}(\mat{D}) \arr{x}$ can be computed with $p$ multiplications of $\mat{T}_{x}(s)$ by a vector. These multiplications can be done in a fast way using a Multilevel Fast Multipole Method (MLFMM) in a  $O(\Ns \text{log}(\Ns))$ complexity \cite{song1997multilevel,frangi2010application}.
In summary, the computational cost for the multiplications of  $\mat{Z}_{j}^{\mbox{reg}}$ by vectors is $O(Q(N_{\text{iter}} + N_{\text{conv}}) \Nt \Ns \text{log}(\Ns))$ operations overall and $O(Q \Ns \text{log}(\Ns))$ in memory, which is the dominant complexity of the solver.

\section{Numerical results}
\label{sec:Numerical_results}

In the following numerical results we have used a modulated Gaussian plane wave for the excitation 
\begin{equation}
    \vt{e}^\text{inc}(\vt{r}, t) = \uv{p} \me^{-\frac{\tau^2}{2 \sigma^2}} \mbox{cos}(2 \pi f_0 \tau) A_0 \,,
\end{equation}
where $\tau(\vt{r}, t) = t - \uv{k} \cdot \vt{r} / c_0$ is the delay, $\uv{p} = \uv{x}$ is the polarization, $\uv{k} = -\uv{z}$ is the direction of propagation, and $A_0 = 1$ V/m is the peak amplitude. $f_0$ is the central frequency and $\sigma$ is the characteristic time that essentially depends on the frequency bandwidth of the excitation.
Given the function $\vt{e}^\text{inc}$ above, its time primitives $\vt{E}^\text{inc}_-$ and $\vt{E}^\text{inc}_+$ that are equal to $\vt{0}$ when $t$ goes respectively to $-\infty$ and $+\infty$, used in \eqref{eq:primitive_en_t}, are equal to
\begin{align}
    \vt{E}^\text{inc}_-(\vt{r}, t) &= \uv{p} \alpha \Re(\mbox{erfc}(-\beta)) \,, \\
    \vt{E}^\text{inc}_+(\vt{r}, t) &= -\uv{p} \alpha \Re(\mbox{erfc}(\beta)) \,, \\
    \alpha &= \sqrt{\frac{\pi}{2}} \sigma  \me^{-2 \pi^2 f_0^2 \sigma^2} A_0 \,, \\
    \beta &= \frac{\tau + 2 \im \pi f_0 \sigma^2}{\sqrt{2} \sigma} \,,
\end{align}
where $\Re(\mbox{erfc}(\cdot))$ is the real part of the complementary error function.
In the following simulations, we used $Q = 16$ and $\rho = 1 + 10^{-4}$ for the inverse Z transform \eqref{eq:trapezoidal_rules}.
The IRK method used is the 3 stages Radau IIA (fifth order) whose Butcher tableau is
\begin{equation}
\label{eq:butcher_tableau_3stages_radau_IIA}
{\renewcommand{\arraystretch}{1.5}
\begin{tabular}{ c | ccc }
    $\frac{4-\sqrt{6}}{10}$  & $\frac{88-7 \sqrt{6}}{360}$ &  $\frac{296-169  \sqrt{6}}{1800}$ & $\frac{-2+3 \sqrt{6}}{225}$ \\ 
    $\frac{4+ \sqrt{6}}{10}$ & $\frac{296+169 \sqrt{6}}{1800}$ & $\frac{88+7 \sqrt{6}}{360}$ & $\frac{-2-3 \sqrt{6}}{225}$ \\ 
    $1$ & $\frac{16- \sqrt{6}}{36}$ &  $\frac{16+ \sqrt{6}}{36}$ & $\frac{1}{9}$ \\ 
    \hline
      & $\frac{16-\sqrt{6}}{36}$ &  $\frac{16+\sqrt{6}}{36}$ & $\frac{1}{9}$
\end{tabular} 
} \,.
\end{equation}
Also, the length parameter $a$ is fixed to 1m.

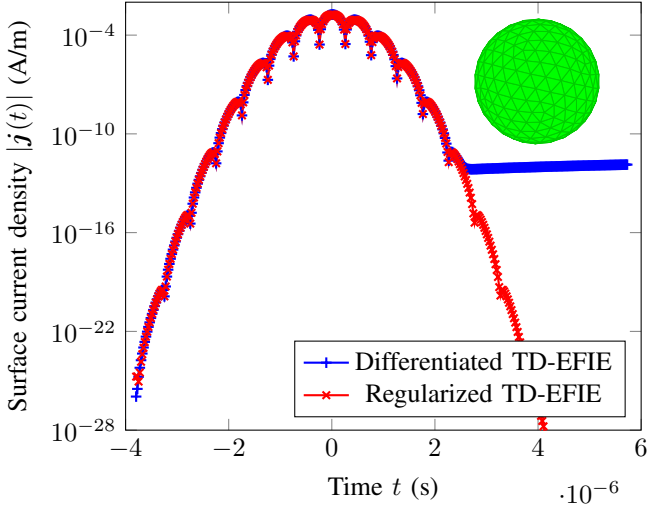
\begin{figure}
    \centering
        \begin{tikzpicture}[scale=1.0]
        \begin{axis}[
                xlabel={Time $t$ (s)},
                ylabel={Surface current density $|\vt{j}(t)|$ (A/m)},
                ymode=log,
                xmin=-4e-6, xmax=6e-6,
                ymax=1e-2, ymin=1e-28,
                legend pos=south east,
                ]
        \addplot+[color=blue,mark=+,thick,each nth point=1] table [x=time, y=currentNoPrecond]{figures/fig1.txt};
        \addlegendentry{Differentiated TD-EFIE}
        \addplot+[color=red,mark=x,thick,each nth point=1] table [x=time, y=currentPrecond]{figures/fig1.txt};
        \addlegendentry{Regularized TD-EFIE}
        \end{axis}
        \begin{axis}[axis equal,scale=0.75, axis lines=none,shift={(2.9 cm,2.5 cm)}]
        \addplot3[patch, colormap name=unicolormapgreen,  patch table= {figures/meshes/sphere_faces.txt}] file {figures/meshes/sphere_vertices.txt};
        \end{axis}
    \end{tikzpicture}
    \caption{Norm of the current density as a function of time on a unit sphere.}
    \label{fig:current_fn_t}
\end{figure}

The first experiment demonstrates the absence of DC instability in the solution. We have used a unit sphere with $\Ns = 750$ edges. There are $\Nt = 400$ time steps between $t_0 = -10 \sigma$ and $t_{\text{end}} = 15 \sigma$ ($\Delta t = 23.9$ ns, $\sigma = 382$ ns, $f_0 = 1$ MHz).
Fig. \ref{fig:current_fn_t} shows the norm of the current in the point (0.45m, 0.88m, 0.06m).
It can be observed that the regularized formulation does not suffer from the DC instability as the solution keeps vanishing where it grows exponentially for the differentiated TD-EFIE. It has been tested numerically that this spurious current verifies \eqref{eq:recurrence_dc_current_diff_tdefie}.

\begin{figure}
    \centering
        \begin{tikzpicture}[scale=1.0]
        \begin{axis}[
                xlabel={Time $t$ (s)},
                ylabel={Surface current density $|\vt{j}(t)|$ (A/m)},
                ymode=log,
                xmin=-4e-6, xmax=6e-6,
                ymax=1e-2, ymin=1e-28,
                legend pos=south east,
                ]
        \addplot+[color=blue,mark=+,thick,each nth point=1] table [x=time, y=currentNoPrecond]{figures/fig2.txt};
        \addlegendentry{Differentiated TD-EFIE}
        \addplot+[color=red,mark=x,thick,each nth point=1] table [x=time, y=currentPrecond]{figures/fig2.txt};
        \addlegendentry{Regularized TD-EFIE}
        \end{axis}
        \begin{axis}[axis equal,scale=0.45, axis lines=none,shift={(3.9 cm,3.5 cm)}]
        \addplot3[patch, colormap name=unicolormapgreen,  patch table= {figures/meshes/torus_faces.txt}] file {figures/meshes/torus_vertices.txt};
        \end{axis}
    \end{tikzpicture}
    \caption{Norm of the current density as a function of time on a torus.}
    \label{fig:torus_current_fn_t}
\end{figure}
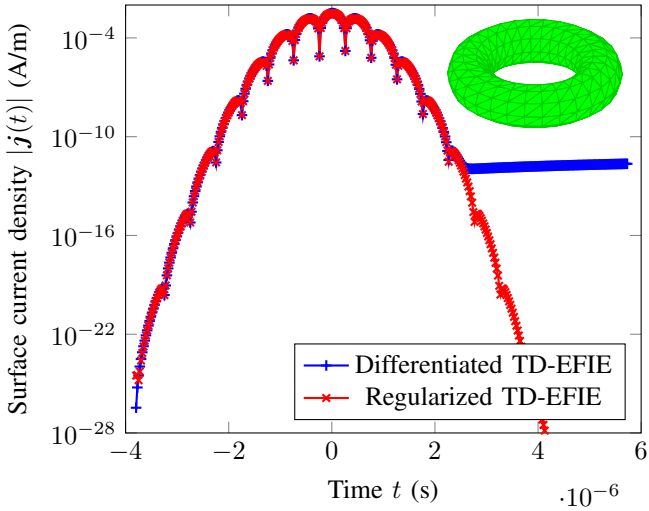

The formulation works perfectly in the case of multiply connected geometries. In this second experiment, we have done the simulation with a torus ($\Ns = 900$ edges) those inner radius is 0.5m and the outer radius is 1m, with the same parameters as the previous example. Fig. \ref{fig:torus_current_fn_t} shows the norm of the current in the point (0.97m, 0.14m, 0.05m).
Again, we can see that the regularized formulation provides the same correct result as the differentiated TD-EFIE but does not suffer from the DC instability.

\begin{figure}
         \includegraphics{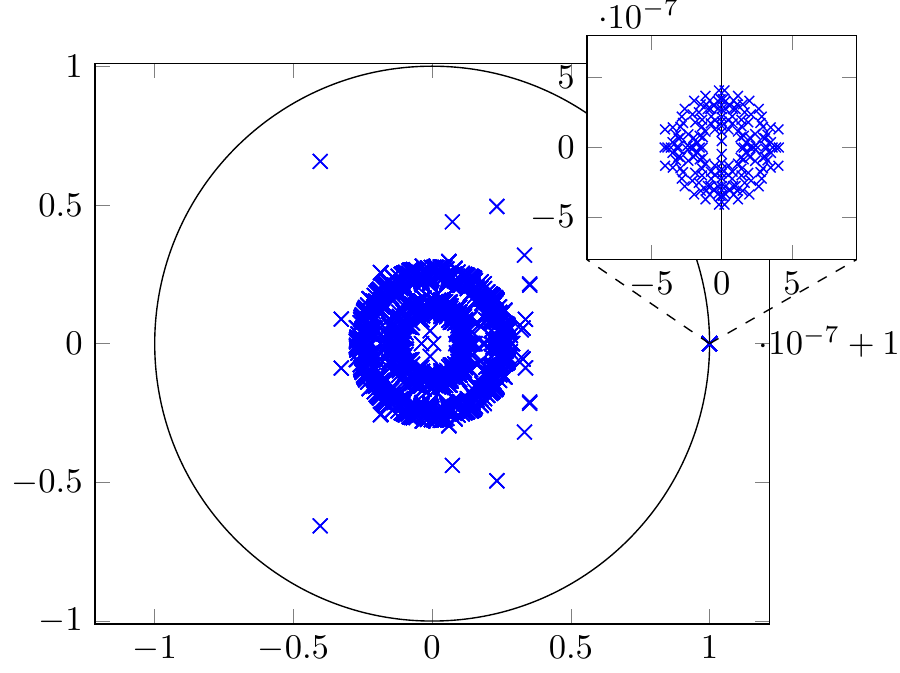}
        \caption{Polynomial eigenvalues of the non-regularized (differentiated) TD-EFIE system showing a cluster at 1.}
        \label{fig:polyeig_nonprecond}
\end{figure}

\begin{figure}
        \includegraphics{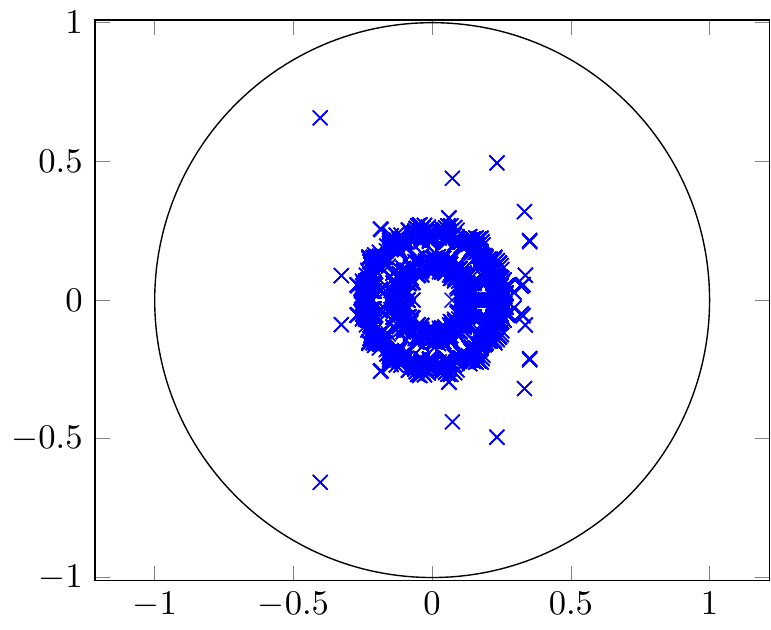}
        \caption{Polynomial eigenvalues of the regularized (preconditioned) TD-EFIE system (there is no cluster at 1).}
        \label{fig:polyeig_precond}
\end{figure}

Another way to observe the effect of the preconditioner on the DC instability is to plot the polynomial eigenvalues associated to the sequence $(\mat{Z}_{j})$ \cite{andriulli2009time}.
The polynomial eigenvalues are the eigenvalues of the companion matrix that corresponds to a MOT in the absence of excitation. Therefore the formulation is stable only if all the eigenvalues are inside the unit circle because any current will vanish exponentially in the absence of excitation.
On the contrary, if there are eigenvalues outside of the unit circle, the error in the solution will grow exponentially. The figures \ref{fig:polyeig_nonprecond} and \ref{fig:polyeig_precond} plot the polynomial eigenvalues associated to the sequences of interaction matrices discretized with the IRK scheme in the cases of the time differentiated TD-EFIE (large time step and DC unstable) and the regularized TD-EFIE (this work that is stable).
A unit sphere with $\Ns = 270$ unknowns is used to approximate the solution and the time step is set to $\Delta t = 5.31$ ns.
Also $N = 16$ matrices are used in each sequences. The regularization removes the cluster of eigenvalues in 1 that is indeed the source of the DC instability.
A zoom on the cluster reveals that some of the eigenvalues have an absolute value greater than 1.

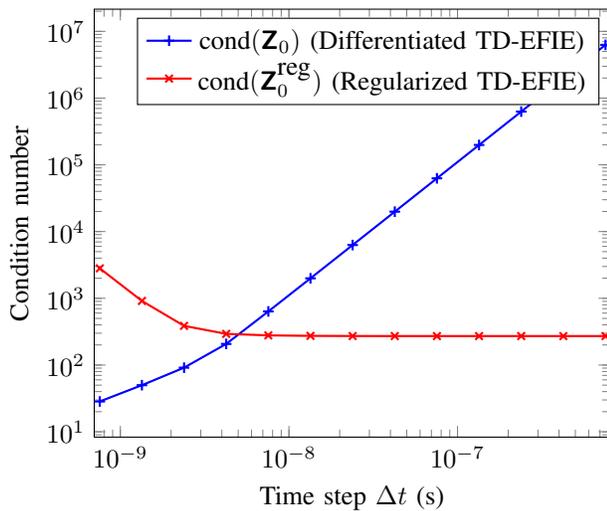
\begin{figure}
    \centering
        \begin{tikzpicture}[scale=1.0]
        \begin{axis}[
                xlabel={Time step $\Delta t$ (s)},
                ylabel={Condition number},
                xmode=log, ymode=log,
                xmin=7e-10,xmax=8e-7,
                ]
        \addplot+[color=blue,mark=+,thick,each nth point=1] table [x=dt, y=condZ0NoPrecond]{figures/fig5.txt};
        \addlegendentry{$\mbox{cond}(\mat{Z}_0)$ (Differentiated TD-EFIE)}
        \addplot+[color=red,mark=x,thick,each nth point=1] table [x=dt, y=condZ0Precond]{figures/fig5.txt};
        \addlegendentry{$\mbox{cond}(\mat{Z}_0^{\mbox{reg}})$ (Regularized TD-EFIE)}
        \end{axis}
    \end{tikzpicture}
    \caption{Condition number of the system as a function of the time step on a unit sphere.}
    \label{fig:cond_fn_dt}
\end{figure}

The low frequency stabilization is demonstrated numerically by computing the condition numbers of $\mat{Z}_0$ for the differentiated TD-EFIE and $\mat{Z}_0^{\mbox{reg}}$ for the stable TD-EFIE for an increasing time step $\Delta t$ on a unit sphere ($\Ns = 750$ as in Fig. \ref{fig:current_fn_t}). It can be seen on Fig. \ref{fig:cond_fn_dt} that there is a quadratic growth of the condition number for the non preconditioned TD-EFIE whereas it is tends to a constant for the regularized formulation. Note that in the small time step limit, a growth of the condition number can be observed. The regularization introduced here is based on the large time step behavior of the equation and does not optimize the conditioning for all time steps.

\section{Conclusion}

We have presented a regularized TD-EFIE that uses implicit Runge-Kutta methods for its time discretization. In particular, the new equation is stable at low frequencies and it does not suffer from the DC instability. The quasi-Helmholtz projectors enable the separation of the two Helmholtz components that can be independently rescaled. On one hand, the time derivative that allows the existence of the spurious static solenoidal current responsible of the DC instability is removed. On the other hand, the ill-scaling of the linear system that is solved is removed which results in a well conditioned system. 




\bibliographystyle{IEEEtran}
\bibliography{ref.bib}

\begin{IEEEbiography}[{\includegraphics[width=1in,height=1.25in,clip,keepaspectratio]{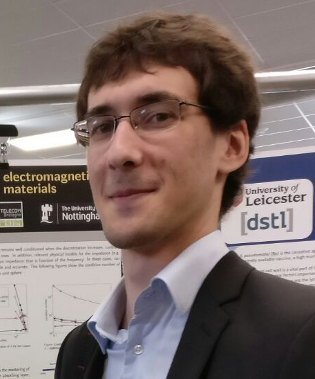}}]{Alexandre D\'ely}
 (S'16)  received the M.Sc. Eng. degree
from the \'Ecole Nationale Sup\'erieure des T\'el\'ecommunications de Bretagne (T\'el\'ecom Bretagne),
France, in 2015. He received the Ph.D.
degree from the \'Ecole Nationale Sup\'erieure Mines-T\'el\'ecom Atlantique (IMT Atlantique), France, and from the University of Nottingham, United Kingdom, in 2019.

He is currently working at Politecnico di Torino, Turin, Italy.
His research focusses on preconditioned and fast solution of boundary element methods, frequency domain and time domain integral equations.
\end{IEEEbiography}

\begin{IEEEbiography}[{\includegraphics[width=1in,height=1.25in,clip,keepaspectratio]{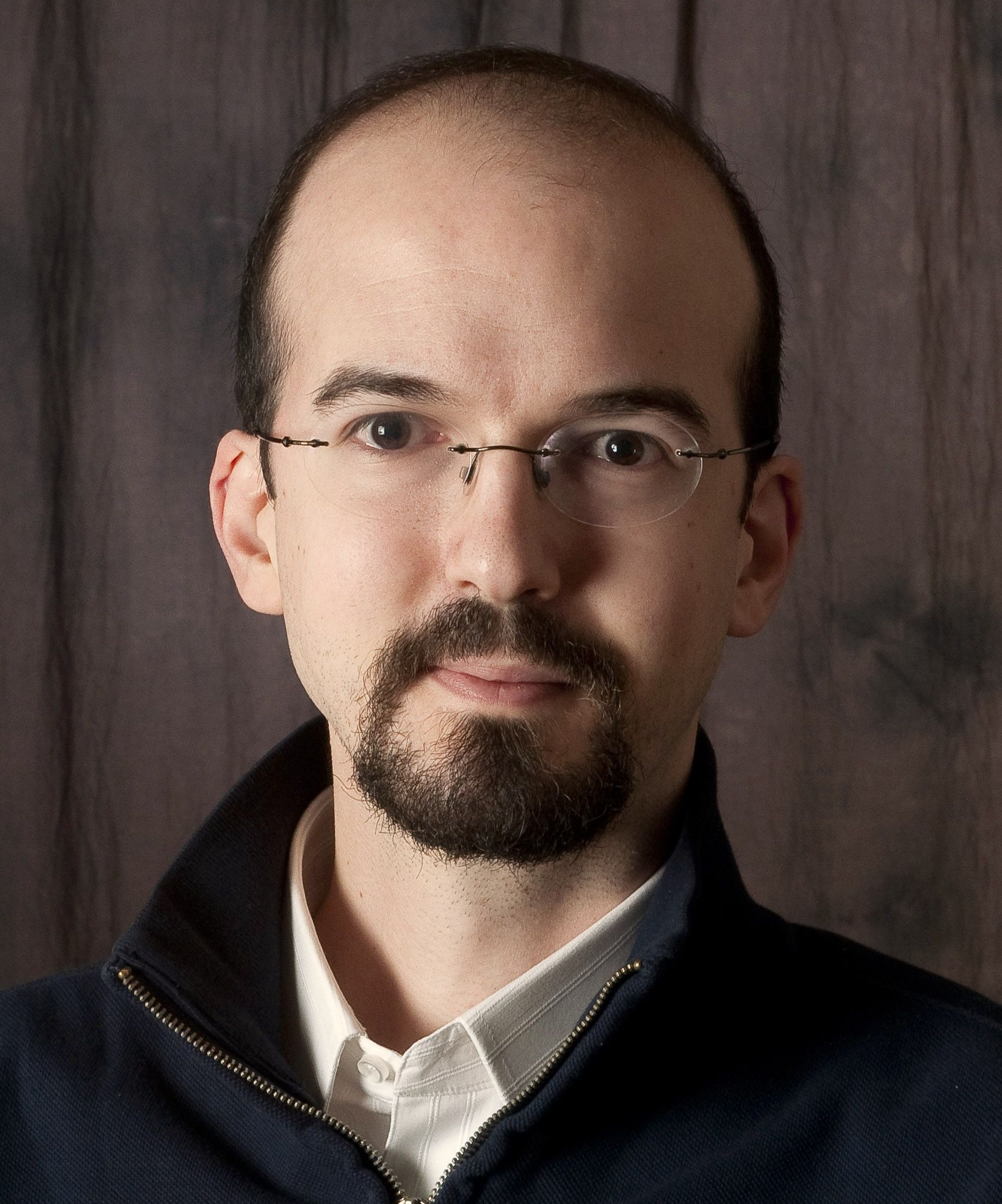}}]{Francesco P. Andriulli}
(S'05-M'09-SM'11) received the Laurea degree in electrical engineering from the Politecnico di Torino, Turin, Italy, in 2004, the M.Sc. degree in electrical engineering and computer science from the University of Illinois at Chicago, Chicago, IL, USA, in 2004, and the Ph.D. degree in electrical engineering from the University of Michigan at Ann Arbor, Ann Arbor, MI, USA, in 2008.

 From 2008 to 2010 he was a Research Associate with the Politecnico di Torino. From 2010 to 2017 he was an Associate Professor (2010-2014) and then Full Professor with the \'Ecole Nationale Sup\'erieure Mines-T\'el\'ecom Atlantique (IMT Atlantique, previously ENST Bretagne), Brest, France. Since 2017 he has been a Full Professor with the Politecnico di Torino, Turin, Italy. 
His research interests are in computational electromagnetics with focus on frequency- and time-domain integral equation solvers, well-conditioned formulations, fast solvers, low-frequency electromagnetic analyses, and modeling techniques for antennas, wireless components, microwave circuits, and biomedical applications with a special focus on Brain Imaging.

He was the recipient of the best student paper award at the 2007 URSI North American Radio Science Meeting.  He received the first place prize of the student paper context of the 2008 IEEE Antennas and Propagation Society International Symposium. He was the recipient of the 2009 RMTG Award for junior researchers and was awarded two URSI Young Scientist Awards at the International Symposium on Electromagnetic Theory in 2010 and 2013 where he was also awarded the second prize in the best paper contest. He also received the 2015 ICEAA IEEE-APWC Best Paper Award. In addition, he co-authored with his students and collaborators other three first prize conference papers (EMTS 2016, URSI-DE Meeting 2014, ICEAA 2009), a second prize conference paper (URSI GASS 2014), a third prize conference paper (IEEE–APS 2018), two honorable mention conference papers (ICEAA 2011, URSI/IEEE–APS 2013) and other three finalist conference papers (URSI/IEEE-APS 2012, URSI/IEEE-APS 2007, URSI/IEEE-APS 2006). Moreover, he received the 2014 IEEE AP-S Donald G. Dudley Jr. Undergraduate Teaching Award, the triennium 2014-2016 URSI Issac Koga Gold Medal, and the 2015 L. B. Felsen Award for Excellence in Electrodynamics. 

Dr. Andriulli is a member of Eta Kappa Nu, Tau Beta Pi, Phi Kappa Phi, and of the International Union of Radio Science (URSI). He is the Editor-in-Chief of the IEEE Antennas and Propagation Magazine, he serves as a Track Editor for the IEEE Transactions on Antennas and Propagation, and as an Associate Editor for the IEEE Antennas and Wireless Propagation Letters, IEEE Access, URSI Radio Science Letters and IET-MAP.

\end{IEEEbiography}

\begin{IEEEbiography}[{\includegraphics[width=1in,height=1.25in,clip,keepaspectratio]{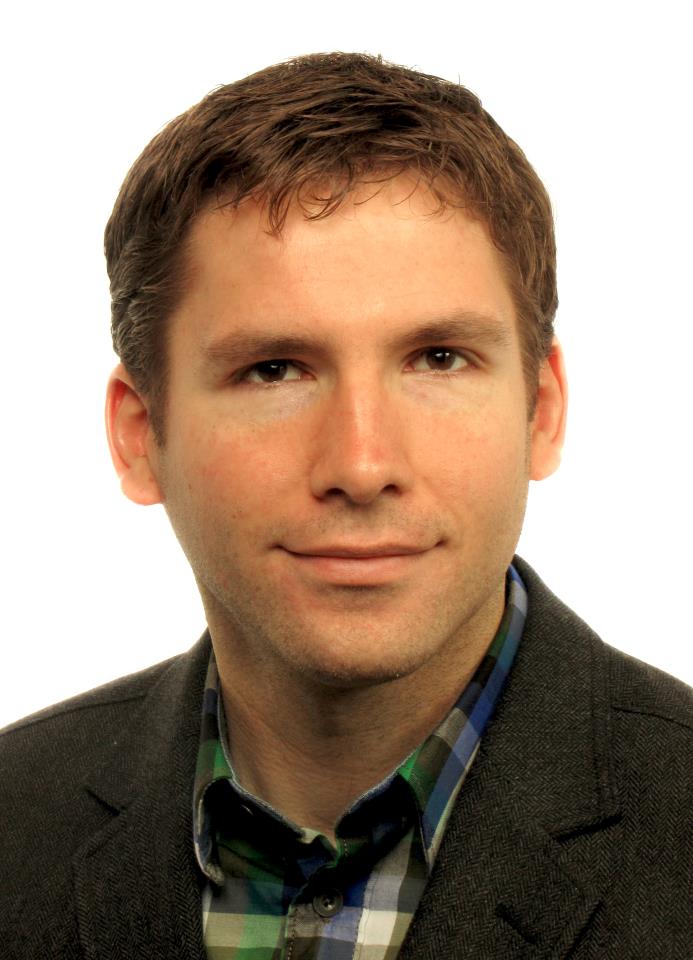}}]{Kristof Cools}
 received the MEng. degree in applied physics engineering and the Ph.D. degree from Ghent University, Ghent, Belgium, in 2004 and 2008, respectively.
 
In 2018, Dr. Cools joined the Department of Industrial and Applied Mathematics at TU Delft. His research interests include the spectral properties of the boundary integral operators of electromagnetics, stable and accurate discretization schemes for frequency and time domain boundary element methods, domain decomposition techniques, and on the implementations of algorithms from computational physics for high-performance computing.

Dr. Cools was the recipient of the Young Scientist Best Paper Award at the International Conference on Electromagnetics and Advanced Applications in 2008. He serves as an Associate Editor for the IEEE Antennas and Propagation Magazine.
\end{IEEEbiography}

\end{document}